\documentclass{aa}
\usepackage{graphics}
\usepackage{rotating}

\def\bv0g{$(B-V)_{0,g}$~}

\def\v5{V$_{0.05}$}
\def\d9{$\Delta t_9$}

\def\ut#1{\mathop{\vtop{\ialign{##\crcr
    $\hfil\displaystyle{#1}\hfil$\crcr\noalign
    {\kern1pt\nointerlineskip}\hbox{$\hfil\sim\hfil$}\crcr
    \noalign{\kern1pt}}}}}
\def\lappreq{\ut {<}}

\begin{document}
\newcommand{\mcol}[3]{\multicolumn{#1}{#2}{#3} }
\newcommand{\struut}{\rule[-2ex]{0ex}{5.2ex}}
\newcommand{\struutup}{\rule{0ex}{3.2ex}}
\newcommand{\struutdown}{\rule[-2ex]{0ex}{2ex}}
\setcounter{equation}{20}

\thesaurus{    11.09.1;
               11.19.5;
               08.06.2;
               03.13.6}

\title{Large scale star formation in galaxies.
I. The spirals NGC 7217, NGC 1058 and UGC 12732
}
\subtitle{Young Star Groupings in Spirals}

\author{P. Battinelli$^1$, R. Capuzzo--Dolcetta$^2$, P.W. Hodge$^3$,
A. Vicari$^1,^2$, T.K. Wyder$^3$}

\institute {$^1$Osservatorio Astronomico di Roma,
Viale del Parco Mellini 84, I-00136 Roma - Italy \\
$^2$Istituto Astronomico, Universit\`a La Sapienza,
via G.M. Lancisi 29, I-00161 Roma - Italy\\
$^3$Astronomy, Department, Box 351580, University of Washington, Seattle,
WA 98195-1580, USA\\
}
\offprints{battinel@oarhp1.rm.astro.it; dolcetta@axrma.uniroma1.it}

\date{Received; accepted}

\authorrunning {Battinelli, Capuzzo-Dolcetta et al.}
\titlerunning{Large scale star formation un galaxies}
\maketitle

\begin{abstract} We present multiband photometric observations of
three spiral galaxies selected in a sample suited for the study  of
young stellar groupings and their relationship with the parent galaxy
and the galactic environment.  Star forming regions have been
identified using an objective technique based on a multivariate
statistical analysis. Maps of young star groupings are given for each
galaxy. The luminosity functions of the young star group populations
show a remarkable similarity with a slope in the range $-1.52$ to
$-1.33$ range. The size distributions peak around the {\it classical}
100pc value of the Local Group associations for two out of the three
galaxies.  NGC 1058 shows smaller associations (peak at $\sim
50$pc). The total number of young groups per unit B absolute
luminosity of the galaxy is significantly greater in UGC 12732. The
activity of star formation is in all three galaxies clearly stronger
in the central regions.
  
\keywords{Galaxies: individual: NGC 7217, NGC 1058, UGC 12732;
Galaxies: stellar content; Stars: formation; Methods: statistical}

\end{abstract}

\section{Introduction}
A few recent papers have pointed
out the importance of the determination of correlations among star forming
region characteristics (size, luminosity, etc.)  with those of the parent
galaxy to gain insights on star formation processes and modes.  For
instance, Elmegreen et
al.  (1994, 1996) suggest the existence of a linear scaling between the
largest
complex size in a galaxy and the size of the galaxy itself. Moreover,
 Elmegreen and
Salzer (1999) find that the largest complex size in each galaxy (in their
sample of 11 galaxies) scales with the square root of the total galaxy
B--luminosity.  These scaling laws are  consistent with
 the largest complex size being equal to 1/2 of the
characteristic Jeans length in the ambient gas, if the gas velocity
dispersion
is always the same in a marginally stable ISM.  In this picture, the
time--scale of star formation is longer (about 50 Myr) for larger galaxies
than
for smaller ones when the size difference is a factor of 10.  This means that
in large galaxies star forming regions should contain both evolved stars and
young OB stars, while in smaller galaxies only OB stars are expected in such
regions.  As a consequence, a correlation between the average age (or
colour) of
the star forming regions and galaxy luminosity is expected.  A related
prediction is that star forming regions in small galaxies should be much
brighter,
but not much larger relative to the host galaxy than those in giant spirals.
It has been clearly shown by Hodge (1986) that any
astrophysical considerations deduced from the collection of heterogeneous
data
should be carefully handled.  Indeed, many biases are introduced by the
difference in the quality of observational data as well as by some degree of
subjectivity unavoidably present in any {\it by--eye} identification of star
groupings.  In order to minimize the uncertainties just described, it
would be
worth dealing with a homogeneous set of observational data (taken with the
same
telescope and equipment) and using an objective method for identifying
blue star
groupings and determining their main properties. Although OB associations
in Local Group galaxies are important to study, it is not possible to derive
significant correlations among star-forming region properties and their
parent
galaxy characteristics using only Local Group galaxies due to the relative
scarcity of galaxies in the Local Group.
For the reasons and
purposes discussed above we decided to start a project which consists of:
{\it i)} the collection of multiband photometric data for a large set of
galaxies beyond the Local Group; {\it ii)} their treatment with the cluster
analysis algorithm described in Adanti et al. (1994). Some preliminary
results are
presented in Wyder et al. (1998).

This paper is the first of a series in which we will present and
discuss  our determination of the young stellar groupings (hereafter
YSGs) in various spirals, chosen among almost face--on galaxies with a
radial velocity limit that corresponds to a moderate distance
($\lappreq 16$ Mpc) and small enough in angular size to be included in
a single CCD frame.  A relevant by--product of this work is an
accurate list of YSG targets to be investigated with higher resolution
observations (e.g. HST--ACS).

We can note that the
philosophy of the present project is the same followed by Bresolin et al.
(1998), who analysed HST data of several nearby galaxies with an automatic
technique suited just for resolved stellar fields.

\section{Observations and data reduction}

Data for the galaxies NGC 7217, NGC 1058 and UGC 12732 were obtained
with the 3.5m telescope at Apache Point
Observatory\footnote{APO is privately owned and operated by the Astrophysical
Research Consortium (ARC), consisting of the University of Chicago, the
Institute for Advanced Study, Johns Hopkins University, New Mexico State
University, Princeton University, the University of Washington and Washington
State University.} using SPIcam, a camera which consists of a
$1024\times1024$
backside illuminated and thinned SITe CCD with read-out noise of 5.7
electrons/pixel and gain of 3.4 electrons/ADU. The image scale is
$0.28\arcsec$/pixel, yielding a field of view of $4.8\arcmin \times
4.8\arcmin$.
UGC 12732 and NGC 7217 were observed in the U,
B, V and I bands on the night of 1997 November 24 while U, B, V and R
images of
NGC 1058 were obtained on 1998 November 21. Both nights were photometric
during the time when the observations were taken, as judged from an all sky
IR camera at the observatory and the standard star observations described
below.

Exposure times for all three galaxies were
1200 s in U and 600 s in each of the remaining filters.
The UGC 12732 and NGC 7217 images consisted of single exposures
while the total exposure time in each filter for the NGC 1058 images
was divided among three exposures. Between these individual exposures
of NGC 1058, the telescope was offset several arcseconds in RA and Dec
to assist in the removal of defects in the CCD.
The seeing averaged about $1.0\arcsec$ (FWHM) for the UGC
12732 images and about $1.3\arcsec$ for NGC 7217 and NGC 1058.

While the V image of NGC 7217 was being
exposed, the telescope focus changed, producing an out of focus image.
On the night of 1997 August 3 a 300 s V image of NGC 7217 was obtained
under non--photometric conditions.  Comparing aperture magnitudes of
several stars in the field between the August 3 and November 24 images,
we were able to calibrate the in focus V image from August 3 and it is
this image that will be used in the following analysis.
 
All of the images were reduced using standard routines in
IRAF.\footnote{IRAF is distributed by the Association of Universities
for Research in Astronomy, Inc., under contract to the National
Science Foundation.} The reduction included the subtraction of the
bias level determined from the overscan region of the CCD as well as
division by an
average twilight flat field image in each filter. An average of
several bias frames obtained at the end of the night showed no
structure, so an average bias frame was not subtracted from the
data.

In some cases the flat field division resulted in a spatial gradient in the
sky background across the image.
The APO 3.5m has little baffling
which may have lead to some contamination of the flat field frames with
scattered light. This scattered light was most apparent in the V, R and I
band
images where the background sky counts are greatest but is likely present
in the U and B data as well.
For example, after division by the flat field, an overall gradient of 3\%
in the sky in the I band image remained. In addition to the likely
presence of
scattered light in the flat fields, the I band images in particular
have a fringing pattern due to the presence of a strong
night sky line within the I passband.  In the 600 s exposures of UGC 12732
and NGC 7217, the fringes have an amplitude of 200-300 ADU in a
background of 6000-7000 ADU. No correction was applied to correct for
the fringe pattern. For this reason the I band data are subject to
larger uncertainties than for the other bands.

The fractional pixel offsets between all of the exposures of each galaxy
was determined using the positions of several stars in the field. Once
the offsets were known, the images were shifted to the nearest fraction
of a pixel using the task IMSHIFT
in IRAF with linear interpolation.

For the NGC 1058 data, the individual exposures in each filter were averaged
together and cosmic rays removed using the the IRAF task IMCOMBINE.
As the UGC 12732 and NGC 7217 data only consist of a single exposure
in each filter, a different method was used to remove the cosmic rays.
Cosmic rays were removed in these images using the task COSMICRAYS, an IRAF
task that identifies pixels which deviate strongly from the
surrounding pixels and replaces the value in those pixels with the
average of the values in the surrounding pixels. This procedure
eliminated almost all of the cosmic rays, mostly leaving those hits
that affected more than one pixel.

The UBVR observations of NGC 1058 from the night of 1998 November 21 were
calibrated using observations of photometric standard stars from
Landolt (1992). The standard stars range in colour from $-1.1$ to 0.8 in
U$-$B,
from $-0.3$ to 1.1 in B$-$V and from $-0.1$ to 0.6 in V$-$R and
were observed several times throughout
the night between airmasses from 1.1 to 1.6. Instrumental aperture
magnitudes for all of
the standard star observations were extracted using a 7$\arcsec$ radius
aperture radius.

On the night of 1997 November 24, observations of three stars in Selected
Area 92 of Landolt (1992) were used to determine the calibration for
the UGC 12732 and NGC 7217 data. The stars range in colour from 0.2 to 1.3
in U$-$B, 0.7 to 1.1 in B$-$V and 0.8 to 1.2 in V$-$I and were observed
at airmasses ranging from 1.2 to 1.5. As for the 1998 data, instrumental
magnitudes were extracted for each observation of the three standard stars.
Since this standard field does not have any very blue stars,
we have fixed the colour coefficients for the UBV data for
this night to be the same as determined above from the 1998 data
to insure that our calibration is valid for the blue colours typical
of YSGs.

\section{The identification techniques}
   
Since the procedure adopted for the identification and classification of star
groupings in unresolved galaxies is fully described in Adanti et al.
(1994), we
just give here a quick overview of the method.  This technique relies on a
combined application of Principal Component Analysis (PCA) and Cluster
Analysis
(CA) to multicolour data of galaxies.  The availability of various sets of
colours
and fluxes allows every pixel of the galaxy image to be represented as a
point
in the space of the variables (i.e. of fluxes and colours).  PCA and CA result
in an artificial image of the galaxy, where pixels are grouped into classes
according to their relative distances in the space of variables.  Our
methodology has a large range of possible applications, and we, in
particular,
proved its reliability for the determination of star forming regions
through a
detailed test on M 31 (see Adanti et al 1994).  We applied our automatic
identification of YSGs on a set of M 31 images artificially degraded in
resolution to simulate it as it would appear at a distance of 10 Mpc.  The
results were found in good agreement with previously available
identifications
(van den Bergh 1964; Efremov et al. 1987).

In order to
make the whole identification procedure easy to handle and its interpretation
 straightforward, we have implemented a friendly IDL based
interface.  The use of this facility is particularly convenient for
projects --
like the present one -- which require an extensive analysis for a large
set of
galaxies.

The background correction for YSGs is intrinsically a difficult task
to perform, because it is not possible to apply the procedures usually
adopted for individual stars.  YSGs indeed tend to be located in an
environment rich with unresolved structure (dust lanes, stars and
clusters, etc.).  In any case we found that an effective
identification of YSGs requires a subtraction of the underlying large
scale luminosity gradients of the galaxy.  This subtraction makes,
first of all, the individual structures emerge better from the
background. Furthermore, if we do not perform the subtraction before
the application of our CA technique, we found that many of the classes
resulting from the CA algorithm are used to account for  the overall
luminosity profile of the galaxy, thus limiting the classes available
to delineate the structures (e.g. YSGs) that we are interested in.

A reliable estimate of the large
scale galaxy luminosity profile has been obtained as follows:  1) areas
where no
evidence of spiral arms and other structures were selected throughout the
images; 2) for each radial bin the $20^{th}$ percentile of the photon
counting
distribution was determined; 3) an analytical fit to this radial distribution
was performed.  Our adopted choice of a low
level (the $20^{th}$ percentile) in the flux distribution (see point 2) is
justified by the necessity of removing "luminous" structures (e.g. faint,
diffuse spiral arms) incidentally
present in the selected sampling areas.  An a--posteriori check of the
influence
of the percentile level showed that the final identification of YSGs is just
marginally affected:  a variation of the threshold in the $10-30$ $^{th}$
percentile range does not affect the number of identified objects,
while has a negligible influence ($<5\%$) on their sizes.  We found that a
decreasing exponential law well represents the
radial luminosity profiles (in the various bands) for the  galaxies
studied here.  Correlation factors are always greater than $0.97$.
Fig.  1
shows an example of the obtained fitting profile.

\begin{figure}[ht]
\label{pha22}
 \resizebox{\hsize}{!}{\includegraphics[height=0.5 cm,angle=0]{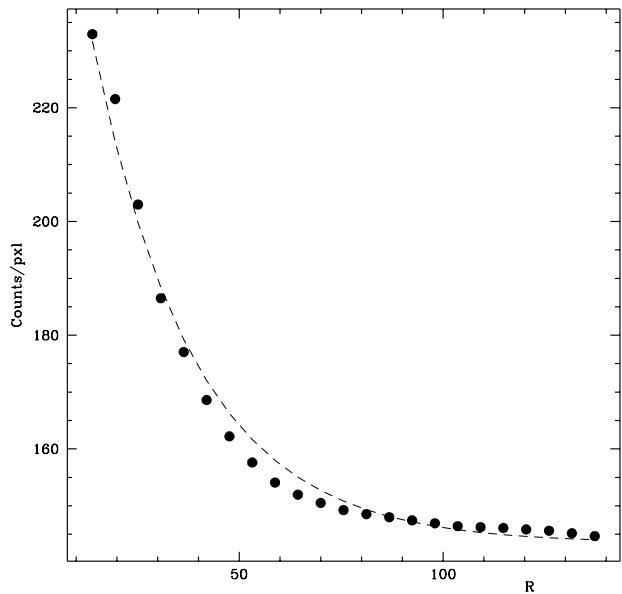}}
 \caption{Background radial profile
(in counts/pixel)  for the galaxy NGC 1058 used in the
identification of YSGs as described in Section 3. The radius R is in pixels. }
\end{figure}

While the above procedure is good enough to improve the identification
of YSGs, it is clearly insufficient to estimate a reliable local
background for each YSG.  Thus, we attempted to determine the local
background in a more sophisticated way:  we selected a rectangular
region surrounding the YSG and fitted a plane to the background
values.  As an estimate of the error we take the standard deviation of
the residual counts after the fitted background subtraction.  This
error can go up to $0.5$ mag (at low luminosities) in all the
photometric bands, rendering colours for YSGs unreliable.  Even for
the luminosities we feel confident only on a statistical use of them
(e.g.  luminosity functions), rather than on individual values. However, these
uncertainties do not affect the identification of YSGs or the
determination of their morphological parameters.

\section{Results}

\subsection { NGC 7217}
The galaxy NGC 7217 is an almost face--on spiral of Sb
type, with no major neighbours (the nearest galaxy is the small irregular NGC
7292, at about 1.3 Mpc distance).  NGC 7217 is 15.5 Mpc distant (assuming
H$_0$=75 km/sec/Mpc) with an integrated B magnitude of 11.17 mag and
isophotal diameter D$_{25}=3.9\arcmin$.  The physical scale of the galaxy
image is 21 pc/pixel.  Images of NGC 7217 show, as already noted by
Verdes--Montenegro et al.  (1995) and Buta et al.  (1995), a
3-ring structure characterized by a high H$_\alpha$ luminosity, suggesting
an active current star formation rate.  In particular, the outermost ring
(about 6 kpc from the galactic centre) contains about $2/3$ of the neutral
hydrogen mass in the galaxy.  Usually, the presence of rings in galaxies is
associated with evidence of bar--like structures (see Buta \& Crocker
1991 and references therein). However, NGC 7217 is remarkably axisymmetric
and no
bar has been detected so far even in the infrared.

The application of our classification algorithm for this galaxy led to
an identification of 149 YSGs, which are shown in Fig.  2b.  We
clearly recover in this figure the multiple ring--like structure.  The
main characteristics of the groups are given in Table 1.  The size, D,
has been defined as the average between the x and y elongations of the
object on the frame.  The size histogram based on the data of Table 1
(see Fig. 3) peaks around 100 pc (average and standard deviation of 130
pc and 84 pc, respectively) similarly to other Local Group
galaxies. If the YSG luminosity was negative after background
subtraction, we entered n.a. (not available) for the integrated B
magnitude and surface brightness $\Sigma_B$ in this table, as well as
in Tables 2 and 3.

\begin{figure}[ht]
\label{pha22}
 \resizebox{\hsize}{!}{\includegraphics[height=0.5 cm,angle=0]{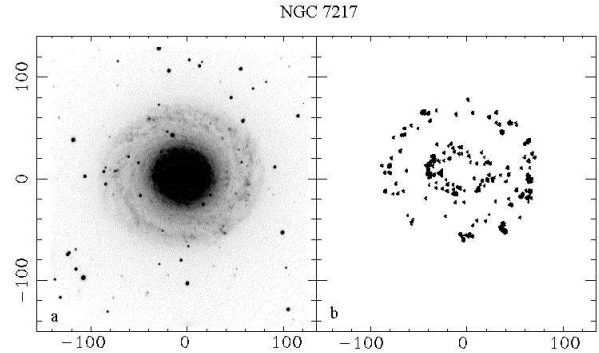}}
 \caption{panel a:  B-band image of NGC 7217 with
North up and East to the left.
 panel b: map of the YSGs identified by our algorithm in NGC 7217.
Coordinates are in arcseconds, with the offset at the galactic centre.}
\end{figure}

We converted the integral magnitudes for each YSG to absolute
luminosities  using our assumed distance.  Luminosities were corrected
for a galactic foreground extinction of A$_B=0.41$ mag, i.e. the
 value listed in the Third Reference Catalogue of Bright Galaxies
(RC3) (de Vaucouleurs et al. 1991).  The resulting differential
luminosity function is shown in Figure 4.  At its bright end the
luminosity function can be approximated as a power law while the
distribution turns over below log$(L_B/L_{B,\odot})=5.0$ due to
incompleteness.  Fitting a power law of the form $dN \propto
L^{\alpha} dL$ to the points brighter than the turnover, we find
$\alpha=-1.46 \pm 0.20$.

In Figure 5, we show the YSG B-band surface brightness averaged over
concentric
circular annuli as a function of the galactocentric radius.  The YSG
positions
were corrected for the galaxy's inclination using the optical minor to major
axis ratio of 0.83 and position angle of 95$^\circ$ from the RC3.  The YSG
B-band luminosity per unit area is related to the high mass star formation
rate.
Figure 5 shows that the peak values of the star formation rate in the two
rings are an order of magnitude higher than in the inter-ring region assuming
that the dust absorption, IMF and YSG age distribution do not vary with
galactocentric distance.

\begin{figure}[ht]
\label{pha22}
 \resizebox{\hsize}{!}{\includegraphics[height=0.5 cm,angle=0]{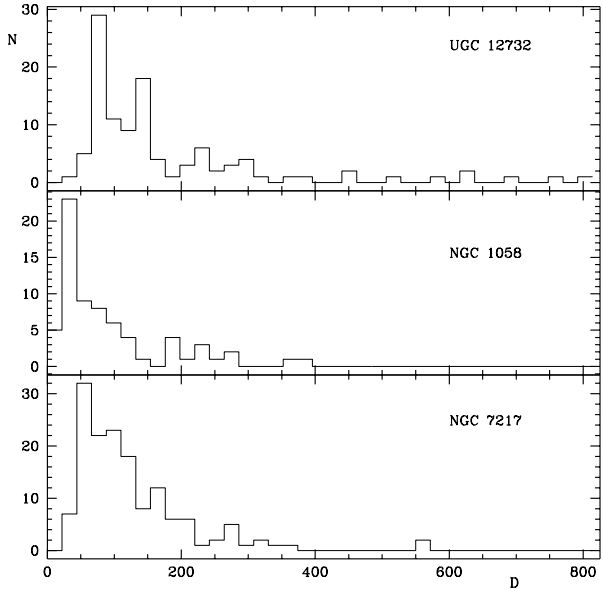}}
 \caption{The size (D, in pc) distribution of the
YSGs in NGC 7217, NGC 1058 and UGC
12732.  The size is defined to be the average of the x and y extents of the
YSG in the image.}
\end{figure}

\begin{figure}[ht]
\label{pha22}
 \resizebox{\hsize}{!}{\includegraphics[height=0.5 cm,angle=0]{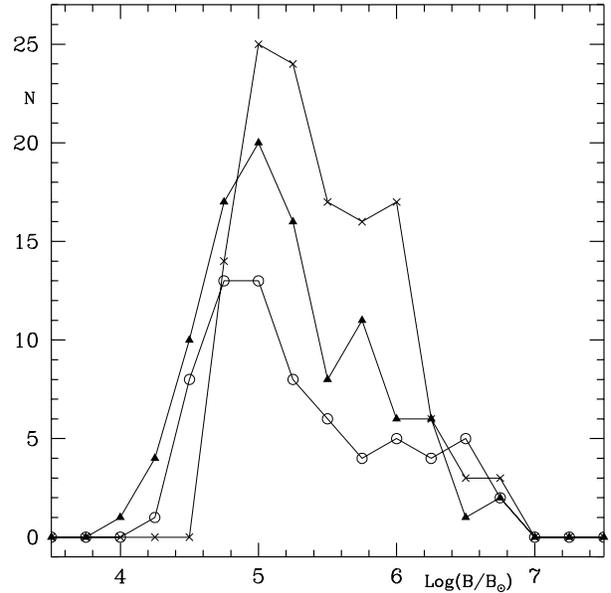}}
 \caption{The B-band differential luminosity
function for the YSGs in NGC
7217 (crosses), NGC 1058 (open circles) and UGC 12732 (black triangles).
In its bright range the luminosity function in all three
galaxies can be fitted by a power law with slopes $-1.46 \pm 0.20$, $-1.33
\pm
0.26$ and $-1.52 \pm 0.25$ in NGC 7217, NGC 1058 and UGC 12732,
respectively.}
\end{figure}

\begin{figure}[ht]
\label{pha22}
 \resizebox{\hsize}{!}{\includegraphics[height=0.5 cm,angle=0]{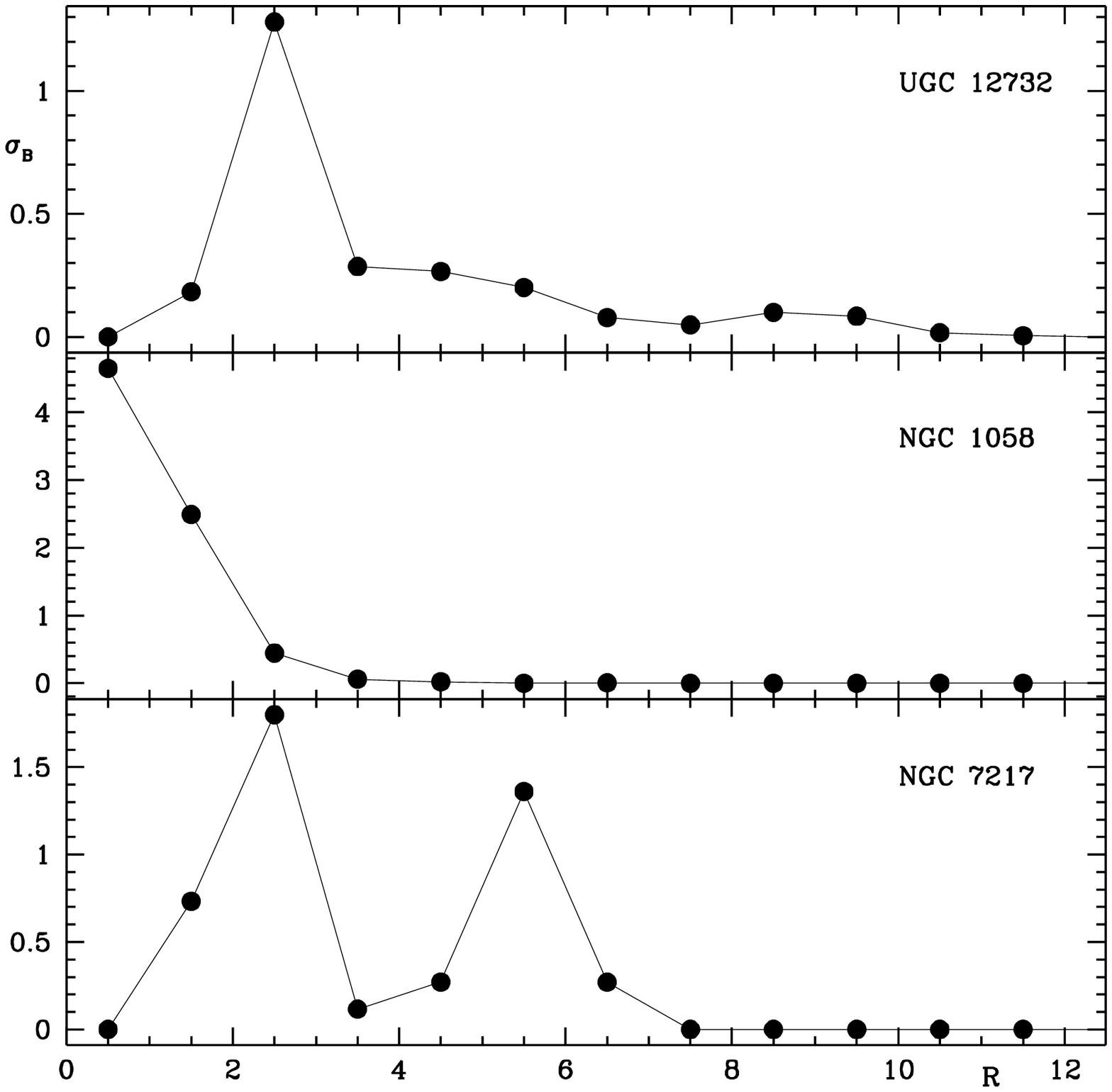}}
 \caption{The surface luminosity density
($\sigma_B$, in $10^6 L_{B,\odot}$ kpc$^{-2}$) of YSGs averaged over
cocentric circular annuli as a function of the galactrocentric radius (R,
in kpc) in NGC
7217, NGC 1058 and UGC 12732.  The YSG positions were corrected for the
galaxy
inclination.}
\end{figure}

\subsection { NGC 1058}

NGC 1058 is a nearly face-on field Sc galaxy with B=11.89 mag and D$_{25}=3.0
\arcmin$ at a distance of 8.4 Mpc.  The physical scale of our CCD images
is 11
pc/pixel.  This galaxy shows a clear knotty structure not organized into
spiral
arms.  Some recent studies by Ferguson et al.  (1998a,b) of its H$\alpha$
regions indicate the presence of a radial metallicity and reddening
gradient in
the sense that inner regions are more metal rich and more reddened, a trend
often observed in spirals.  As for NGC 7217, Table 2 gives all the
relevant data
for the set of 71 YSGs  identified in this galaxy, whose map is shown in
Fig. 6b.

\begin{figure}[ht]
\label{pha22}
 \resizebox{\hsize}{!}{\includegraphics[height=0.5 cm,angle=0]{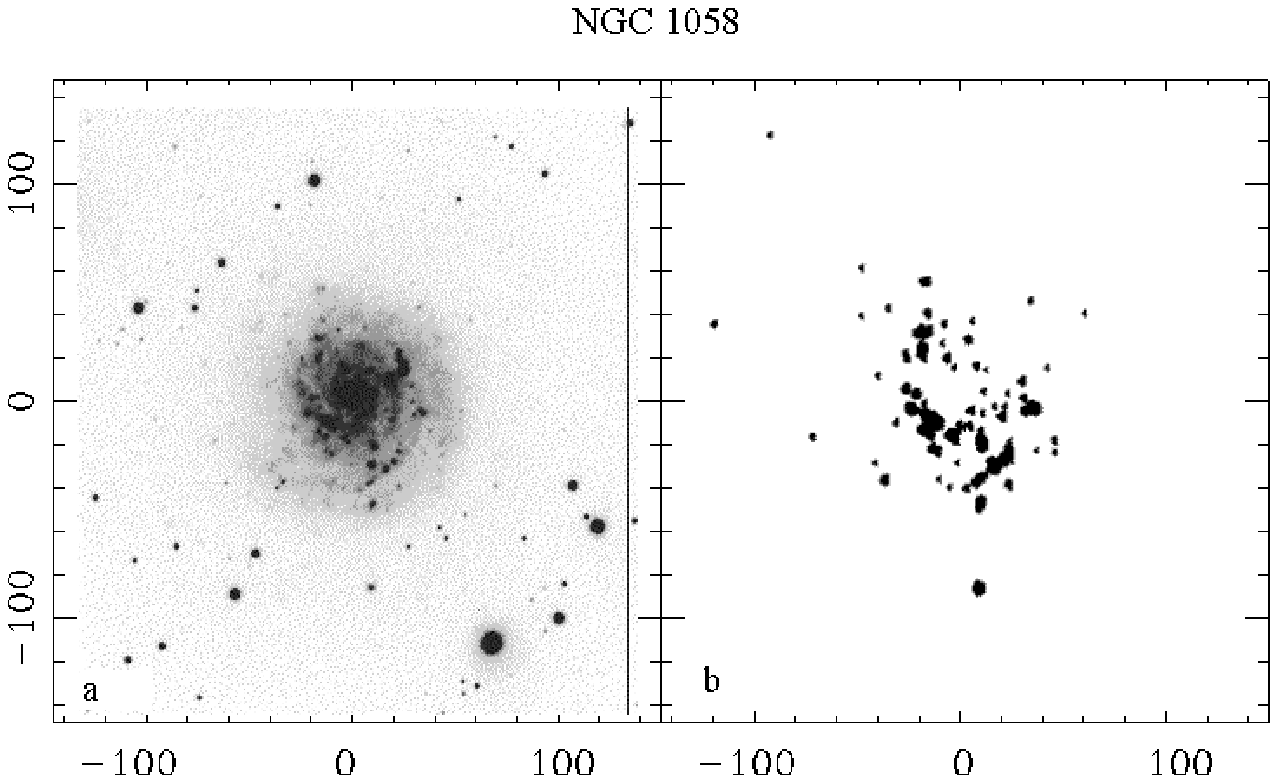}}
 \caption{panel a:  B-band image of NGC 1058 with
North up and East to the left.
 panel b:  Map of the YSGs identified in NCC 1058. Coordinates as in Fig. 2.}
\end{figure}

\begin{figure}[ht]
\label{pha22}
 \resizebox{\hsize}{!}{\includegraphics[height=0.5 cm,angle=0]{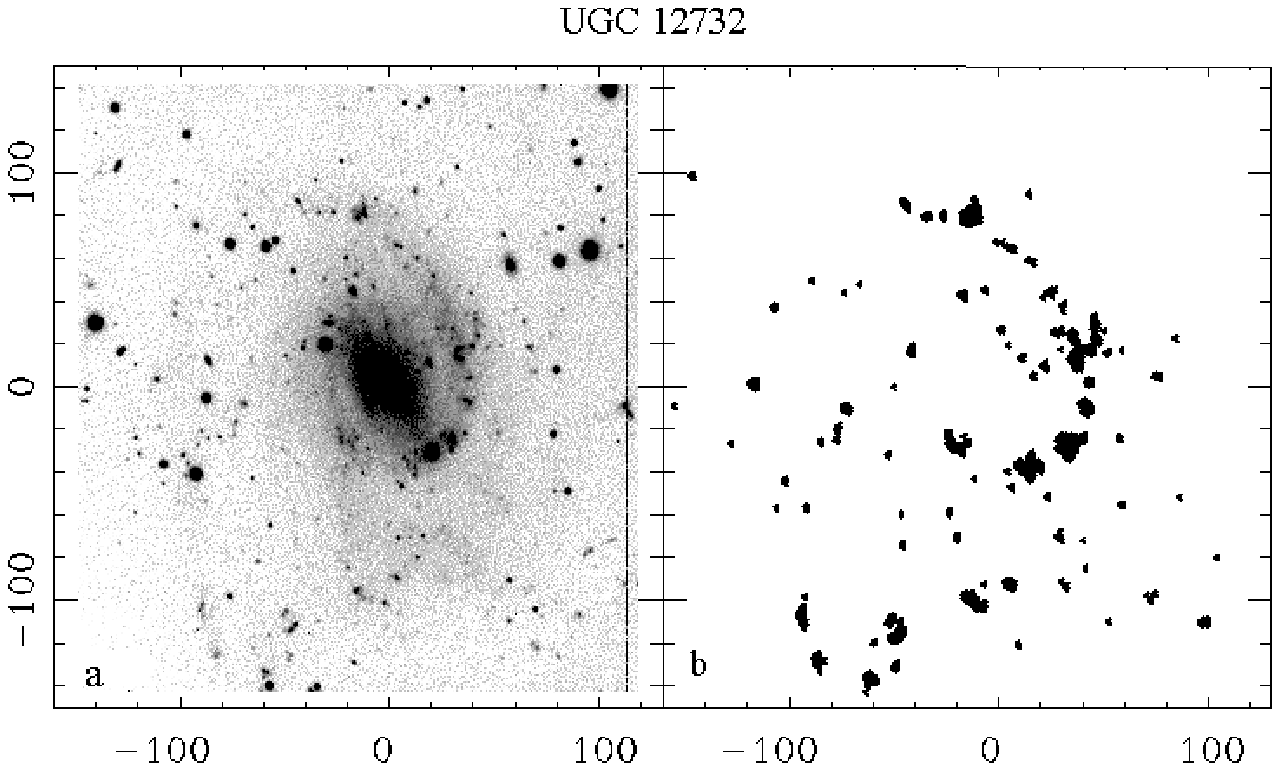}}
 \caption{panel a:  B-band image of UGC 12732 with
North up and East to the left.
 panel b:  Map of the YSGs identified in UCC 12732. Coordinates as in Fig. 2.}
\end{figure}

The size distribution of Fig. 3 peaks at 50 pc with average value and
standard deviation of 104 pc and 109 pc, respectively.
 There is no clear correlation between the size and the galactocentric
distance.
 
In Figure 4 we show the YSG B-band luminosity function; the decreasing
tail corresponds to a slope of $\alpha=-1.33 \pm 0.26$. The B-band
luminosity surface density
of YSGs shown in Figure 5 monotonically decreases,
 falling quickly by
 four orders of magnitude at a galactocentric distance of 3 kpc.

\subsection { UGC 12732}

According to the data in the RC3, the galaxy UGC 12732, is classified
as type Sm with an integrated B magnitude of 13.8 and optical
diameter D$_{25}=3\arcmin$.  This galaxy is at 12.4 Mpc distance which
corresponds to a scale of 17 pc/pixel in our images.  Its nearest
neighbor is the Scd galaxy NGC 7741 which is at a projected distance
of 160 kpc and has a radial velocity that is within 6 km/s of UGC
12732's velocity.

Our images show that the galaxy consists of a smooth low surface brightness
inner region with a single long arm of YSGs starting from SW of the centre
and
circling around to the north side of the galaxy.  In addition there is a
second
much broader and smoother arm extending to the south of the main body of the
galaxy.  There are many other YSGs scattered throughout the image in areas
where
there is no detectable background light from the galaxy itself.

The HI properties of UGC 12732 were recently studied by Schulman et al.
(1997)
based upon data from Arecibo and the VLA. Their data show that
the HI extends throughout the field of our images with a diameter of
$11.6\arcmin$
at an HI level of 10$^{19}$ atoms cm$^{-2}$. This diameter is almost four
times larger than
the optical galaxy. There is a depression in the HI gas coinciding with the
optical centre of the galaxy while the two arms visible in our optical data
are detected in HI as well at a level of $5-6 \times 10^{20}$ atoms
cm$^{-2}$.

We identify 109 young star groups in this galaxy. Their measured properties
are listed in Table 3 while their locations are shown in
Figure 7b.  The size--histogram shown  in Fig. 3 peaks around 90 pc (the
average is 178 pc, the standard deviation
156 pc) and
--again- few large YSGs ($D> 200$ pc) are present.  The more numerous
population
of small YSGs in this galaxy with respect to those in NGC 7217 is the
result of
the joint effects of the lower surface brightness of UGC 12732 and of its
slightly shorter distance.  The dimensions of the YSGs of this galaxy are
similar to those of YSGs in local group galaxies.

The B-band luminosity function of YSGs in UGC 12732 is shown in Figure 4
and has a best-fit power law exponent  $\alpha=-1.52\pm0.25$. The radial
YSG B-band luminosity surface density in Figure 5 peaks at about 2 kpc
radius and falls off smoothly with radius by one and a half orders of
magnitude at a radius of 12 kpc.

\section{Discussion and Conclusions}

This paper presents the first results of an ongoing project whose main
aim is to collect homogeneous data suited to the study of regions of
recent star formation in a variety of galaxies.  We apply a cluster
analysis algorithm to obtain, in a semi--automatic  way, an objective
identification of young star groupings (YSGs) in UBVRI  frames of NGC
7217, NGC 1058 and UGC 12732.

These are the first spirals we observed from a sample (selected from
the RC3) of nearly face--on galaxies, with radial velocities less than
1200 km/sec and small enough to fit in just one CCD frame of the
Apache Point Observatory Spicam. This sample will allow us to check in
a statistically significant way whether correlations among properties
of YSGs and their parent galaxies exist. Such correlations are useful
to understand better how star formation in galaxies proceeds (see
e.g. Elmegreen 1994, 1996).

We have identified 149, 71 and 108 YSGs in NGC 7217, NGC 1058 and UGC
12732, respectively. These numbers correspond to specific frequencies
$S_N\equiv N_t \cdot 10^{0.4(M_B^0+15)}$ equal to $1.0$,  $3.9$,
$19.7$.  If we consider just YSG brighter than the likely completeness
limit ($L_B=10^5L_{B\odot}$), specific frequencies reduce to
$1.0,~3.1$ and $12.9$ for NGC 7217, NGC 1058 and UGC 12732,
respectively.  In any case, the high $S_N$ value for UGC 12732
indicates an overabundance of YSGs in this galaxy; however, since
distances are from radial velocities, peculiar motions may
significantly affect the total luminosity and consequently the $S_N$ value. For
instance, a variation of $50\%$ in the UGC 12732 distance makes its
$S_N$ value comparable to those of the other two galaxies.

As explained in Section 3, determining an accurate background for the
YSGs is an inherently difficult problem since they usually lie in
areas with a lot of structure. This makes the colours of the YSGs
unreliable but the the luminosities can still be used to determine a
statistically reliable luminosity function. We found that the slope of
the bright tails of the YSG  luminosity functions in these three
galaxies are similar, within the errors,  and lie in the range $-1.36
to -1.52$. In Section 4 we also present the size distribution and
surface density of the YSGs versus galactocentric radius: no
significant variation is found. From the radial distribution of our
identified YSGs and their integrated luminosities we could extract
information about differences in star formation rate within each
galaxy and between galaxies. We found a much steeper radial slope for
the SFR in NGC 1058  while in both NGC 7217 and UGC 12732 the
variation of the SFR between the inner and outer galactic regions is
up to a factor of about 10.  The YSG size distributions (which are of
course affected  by the poorly known galactic distances)    are
similar in NGC 7217 and UGC 12732 (peaks around 100 pc), while NGC
1058 shows smaller YSGs (the peak is at $50$ pc). Since the latter
galaxy is much  closer than NGC 7217 and UGC 12732, one could believe
the smaller sizes of YSGs as an artifact of the intrinsic better
spatial resolution (pc/pixels) of the images. A simple test we
performed clearly proves this is not the case: the identification of
YSGs on images of NGC 1058 artificially re-binned to simulate larger
distances shows that the peak of the size distribution is almost
insensitive on a variation up to $40\%$ of the distance. More
specifically, increasing the distance implies a loss of just the
smallest objects (those smeared out by the image re-binning).


\end{document}